# Superradiant Broadband Magneto-electric Arrays Empowered by Meta-learning

Konstantin Grotov[1,=], Anna Mikhailovskaya[1,a),=], Dmytro Vovchuk[1], Dmitry Dobrykh[1], Carsten Rockstuhl[2,3], and Pavel Ginzburg[1]

[1] School of Electrical Engineering, Tel Aviv University, Tel Aviv 69978, Israel

[2] Institute of Theoretical Solid State Physics, Karlsruhe Institute of Technology, 76131 Karlsruhe, Germany

[3] Institute of Nanotechnology, Karlsruhe Institute of Technology, 76344 Eggenstein-Leopoldshafen, Germany

[=] Contributed equally

[a)] Author to whom correspondence should be addressed: anna2@mail.tau.ac.il

**Abstract**: Laws of electrodynamics constrain scattering cross-sections of resonant objects. Nevertheless, a fundamental bound that expresses how larger that scattering cross-section can be is yet to be found. Approaches based on cascading multiple resonances permitted to push the scattering responses of subwavelength structures and to exceed existing estimators, for which the Chu-Harrington criterion is, potentially, the most commonly considered one. The superradiant empirical limit, addressing scattering performances of near-field coupled resonator arrays, was subsequently developed to tighten existing estimates, setting a new bound that prompted efforts to find structures that exceed it. Here, we demonstrate that genetically designed superscattering structures, encompassing arrays of constructively interfering electric and magnetic dipoles, can build enormously high scatting cross-sections exceeding those imposed by existing criteria in electromagnetic theory including the superradiant empirical limit. After undergoing thousands of evolutionary generations, iterating sizes, mutual orientations, and locations of resonators, the structures approach their heuristically maximized performance, which is unlikely to be obtained by a random distribution given more than a billion trials. As an additional practically valuable parameter, the scattering bandwidth also underwent optimization. We demonstrate that flat wavelength-comparable structures can have significant backscattering alongside more than 40% fractional bandwidth. The result demonstrates the fundamental capability to untighten scattering cross-section from bandwidth limitations. New capabilities of genetic optimization algorithms, equipped with fast computational tools and constrained by experimentally obtainable electromagnetic parameters, allow chasing well-accepted traditional bounds, demonstrating ever-seen electromagnetic performances.

*corresponding author

## I. INTRODUCTION

Wave scattering governs multiple universal fundamental phenomena across many disciplines. Electromagnetic scattering is unique due to its enormous practical significance, motivated by wireless applications[1]. Numerous efforts were devoted to assessing the maximally achievable gain, bandwidth, and scattering cross-section, especially considering subwavelength structures [2,3]. However, regardless of its physical size, a dipole scattering cross-section is bounded from above by $\frac{3\pi}{2}\lambda^2$, where $\lambda$ is the resonance wavelength[4]. This number is referred to as the single-channel dipolar limit, which can be further generalized to account for higher-order resonant multipoles, given a structure that supports them[5]. Another tightly related effect of the size reduction is the scattering bandwidth drop, as was addressed by Chu and Harrington and further explored on many occasions[3,6]. A significant objective, having both fundamental and applied relevance, is to identify tighter bounds, linking between footprints and performances of electromagnetic devices. While quite a few superscattering architectures have been demonstrated though without considering bandwidth as an objective[7–12], near-field coupled arrays of resonators were found to have several significant advantages. The hybridization in these near-field coupled structures generally induces higher-order resonant multipoles, which, given a proper design, can be set to resonate at nearly the same frequency. The excitation of these higher-order resonant multipoles builds up a significant scattering cross-section and allows us to exceed the single-channel dipolar limit [13–16]. However, besides design complexity, additional practical constraints emerge in near-field coupled arrays. Examples are a low tolerance to fabrication imperfections and internal material losses of constitutive elements. Highly resonant structures are simply accompanied by a significant near-field accumulation, which probes, in turn, every possible imperfection. Therefore, as good as it is, the near-field accumulation makes the superscattering structures susceptible to the previously mentioned constraints. Nevertheless, recently demonstrated configurations relying on intensive structural optimizations bypassed conventional limits significantly. The superradiant empirical criterion, assessing subwavelength arrays of resonant dipoles, provided a tighter estimate on achievable scattering cross sessions and served as a new guideline in the field[17].

An additional important aspect associated with the size reduction of electromagnetic devices is the operational bandwidth drop. Being typically observed in an overwhelming majority of subwavelength antenna architectures, this aspect also became subject to developing estimating bounds. For example, demonstrating a broadband superdirective antenna remains an open challenge, though quite a few designs were explored, including ceramic resonators[18,19], multilayer designs[20–22], Huygens sources [23–28], high-impedance surface antennas [29], and several others. While non-Foster matching schemes for bandwidth extension were found to be quite promising [30–33], they face stability issues and might possess other limitations[34]. On a related matter, it is worth noting the bounds on matching efficiency *vs.* bandwidth tradeoff (Bode-Fano criterion[35]) with strategies to bypass it with temporal modulation[36]. Consequently, finding new pathways to untighten the link between size, scattering cross-section, and bandwidth has both fundamental and practical significance.

Here, we explore scattering architectures based on flat arrays of near-field coupled electric and magnetic resonators. Meta-learning algorithms are applied to maximize both the backward scattering cross-section and the operational bandwidth, thus untying the performance from Chu-Harrington Q-factor limitations. The analysis of the optimized structure verifies the resonance cascading concept, where several multipolar resonances overlap in the same frequency domain and coherently build up a significant backscattering over an extended bandwidth. The experimental demonstration in the GHz spectral range verifies the superscattering super-bandwidth properties of the structures.

## II. RESULTS

### *The Structure*

In arbitrary-shaped open resonators with broken spatial symmetry, each mode, excited by an incident illumination, can contribute with a multipole mixture to the far-field scattering[37]. As a result, a coherent multipolar content, including relative amplitudes and phases, is to be optimized to grant a broadband superscattering. In case multiple degrees of freedom in a system are optimized simultaneously to maximize a cost function (backward scattering cross section and the bandwidth, in this case), genetic algorithms and meta-learning are among the preferable candidates[38–49]. Since numerous simulations are performed to evaluate the cost function for many different configurations, the computational efficiency of the forward solver plays a key role. While many numerical algorithms for solving electromagnetic scattering exist, surface integral methods applied to curved metal wires enable very fast calculations[50]. The assessment of even quite complex structures can take as few as several seconds on a personal computer with such a method.

The basic configuration considered in this contribution consists of a two-dimensional array of wires and split-ring resonators (SRRs)[51,52]. Short half-wavelength wires support electrical dipole resonances, while symmetrized SRRs primarily interact with the magnetic field (Fig. 1(a)). Magneto-electric coupling[53,54] within the array grants additional flexibility and, as it will be shown, is responsible for elevating performances. The two-dimensional architecture is chosen from the practical constraints of a realization with printed circuit board (PCB) technology. Considering perspective radar applications, the operational frequencies were selected to fall within the 5-11 GHz range. Those parameters dictate the typical sizes of the resonating elements, which are chosen to be bound in 2 cm x 2 cm squares on the substrate. In the following designs, the resonators are centered in the nodes of a rectangular array (Fig. 1(b)). Figure 1(a) demonstrates the basic elements and the degrees of freedom considered in the optimization. Specifically, the wires' length and orientation to the global Cartesian coordinate system (array axes) are kept as variables. In the case of SRRs, the outer radius and the orientational angle are subject to optimization. The number of degrees of freedom is $3N^2$, where N is the number of squares (N = 3, 4, 5 are considered hereafter). For example, the 5 x 5 design already has 75 degrees of freedom in the optimization. Bigger arrays become electromagnetically large and thus will not be considered.

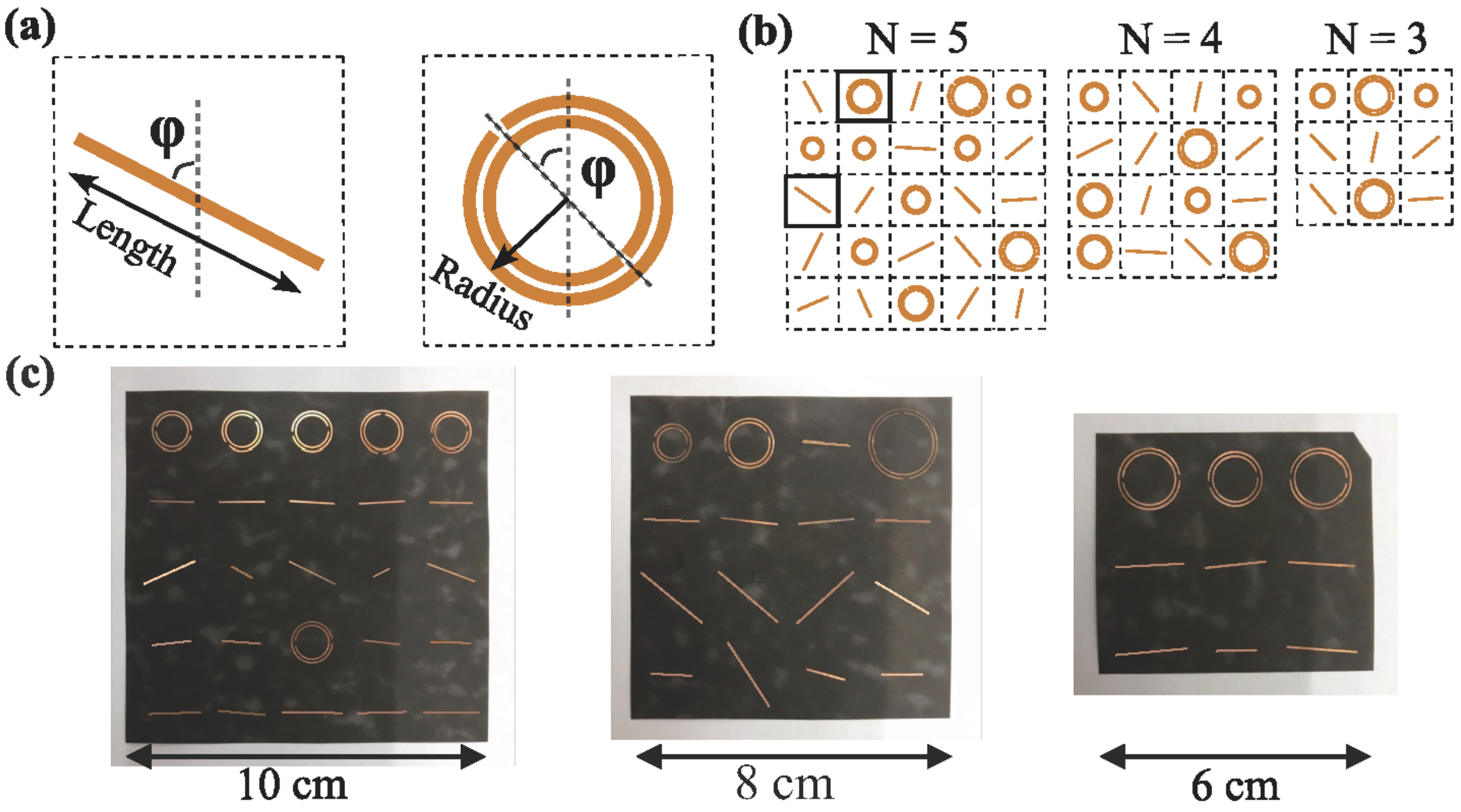

Fig. 1. (a) Wires and split ring resonators are basic elements to populate the arrays. (b) Typical array layouts obtained from the optimization algorithm. (c) Photographs of the fabricated samples (printed circuit board) for the experimental verification (parameters are in Supplementary).

*The Objective Function and Optimization Algorithms*

Optimization algorithms to design electromagnetic structures are typically divided into two categories – those that use gradients of the objective function[55] and those that do not[56]. The choice of a suitable algorithm depends on the details of the problem. In the current situation, the backscattering cross-section spectrum within an extended frequency band, subject to maximization, has quite a complex behavior. In the case of multi-resonant structures, the backscattering cross-section has many local extrema, which can exchange quite fast causing derivative fluctuations. Owing to those harsh properties, most classical optimization methods, such as grid-search, the steepest descent method, or the conjugate gradient method[57], will run into instabilities and might converge to only one out of many local extrema that may or may not be close to the global one. Also, obviously, gradient-based optimizations cannot be applied to the entire problem as there is either a wire or an SRR in the unit cell. Thus, the problem is only partially continuous, and combinational methods are in place.

However, there is another set of entirely different algorithms that do not use gradients during optimization. Specifically, evolutionary methods are based on the principles of natural selection and consist of three key stages: selection, crossover, and mutation[58]. In the first step, a set of individuals is randomly generated by a vector from the search space (a random array of resonators here). Those configurations then undergo a crossover with each other and mutations, forming new individuals, which eventually create a new population. At the end of each epoch, a fixed number of maximally adapted individuals are selected from all new individuals – those who show the largest objective function. At the next optimization step, those selected individuals constitute the initial population. It is worth noting that this approach does not guarantee to find the global maximum of the function, nevertheless, heuristic justifications suggest approaching it. Figure 2(a) summarizes the main optimization steps of the evolutionary algorithm with a pre-defined (static) objective function.

The objective function was to optimize a broadband backscattering cross-section in the 5-11 GHz frequency range over the superradiant limit (number of resonators times the single channel limit[17]). From now on, it will be referred to as the superradiant bandwidth. In the first step, 5-15 equally spaced frequencies (hyperparameters of the evolutionary algorithm) were assigned, at which the backscattering cross-section is evaluated. While a seemingly appealing approach is to tighten the discretization within the frequency band, it leads to elevated computational costs and, more importantly, causes convergence instabilities in the optimization. The latter features emerge from the resonant nature of the superscatterer, whose scattering spectrum is characterized by quite some ripples. Therefore, after calculating the backscattering, an aggregation function was applied – both the mean and the product of the results were assessed. The product function provides stronger penalties to structures with ripples in the spectrum, i.e., those having regions with low scattering, thus, averaging was favored.

The frequencies were manually redistributed to provide more weights to regions with a lower backscattering to improve the bandwidth performances in the next epoch. The logic is depicted in Fig. 2(c). As for the genetic algorithm, a covariance matrix adaptation evolution strategy (CMA-ES) has been used. The maximal number of iterations was manually set to 500, 1000, and 2000 for 3 x 3, 4 x 4, and 5 x 5 structures, respectively. An additional termination constraint was set if less than $10^{-3}$ % changes in the objective function were observed during the last 100 iterations. The initial population was randomly assigned. The population size was chosen to be $3N^2$, where N is the dimensionality of the search space (number of the allowed geometrical degrees of freedom).

To improve the performances and obtain a higher level of automation (to relax the need for manual redistribution of frequencies, where the cost function is sampled (Fig. 2(c)), a meta-learning algorithm was implemented on top (see Fig. 2(b) for the block diagram)[59]. The goal is to use Bayesian optimization to replace the manual redistribution of frequencies within the band by learning approaches. In this endeavor, the spectrum discretization (the number of frequencies) in three clusters was chosen as the parameters (Fig. 2(d)). Overall, the meta-learning algorithm contains eight main parameters - frequencies in each one of the clusters and the bandwidth of each cluster. The fixed parameters of the algorithm are as follows: the number of frequencies within each cluster ranges from 0 to 10, with the frequency step varying from 100 to 500 MHz. The distance between clusters is limited to a 0-1000 MHz range. The number of iterations in the meta-learning algorithm was empirically chosen to be 100. While this number does not guarantee convergence, it dramatically enlarges the search space dimensionality and, thus, leads to elevated performances, as it was juristically shown on other occasions across different disciplines[59]. While comparing the performance of evolutionary optimization and the meta-learning approach is quite difficult owing to manual adjustment in the first case, a rough estimate of several-fold improvement in favor of meta-learning can be claimed.

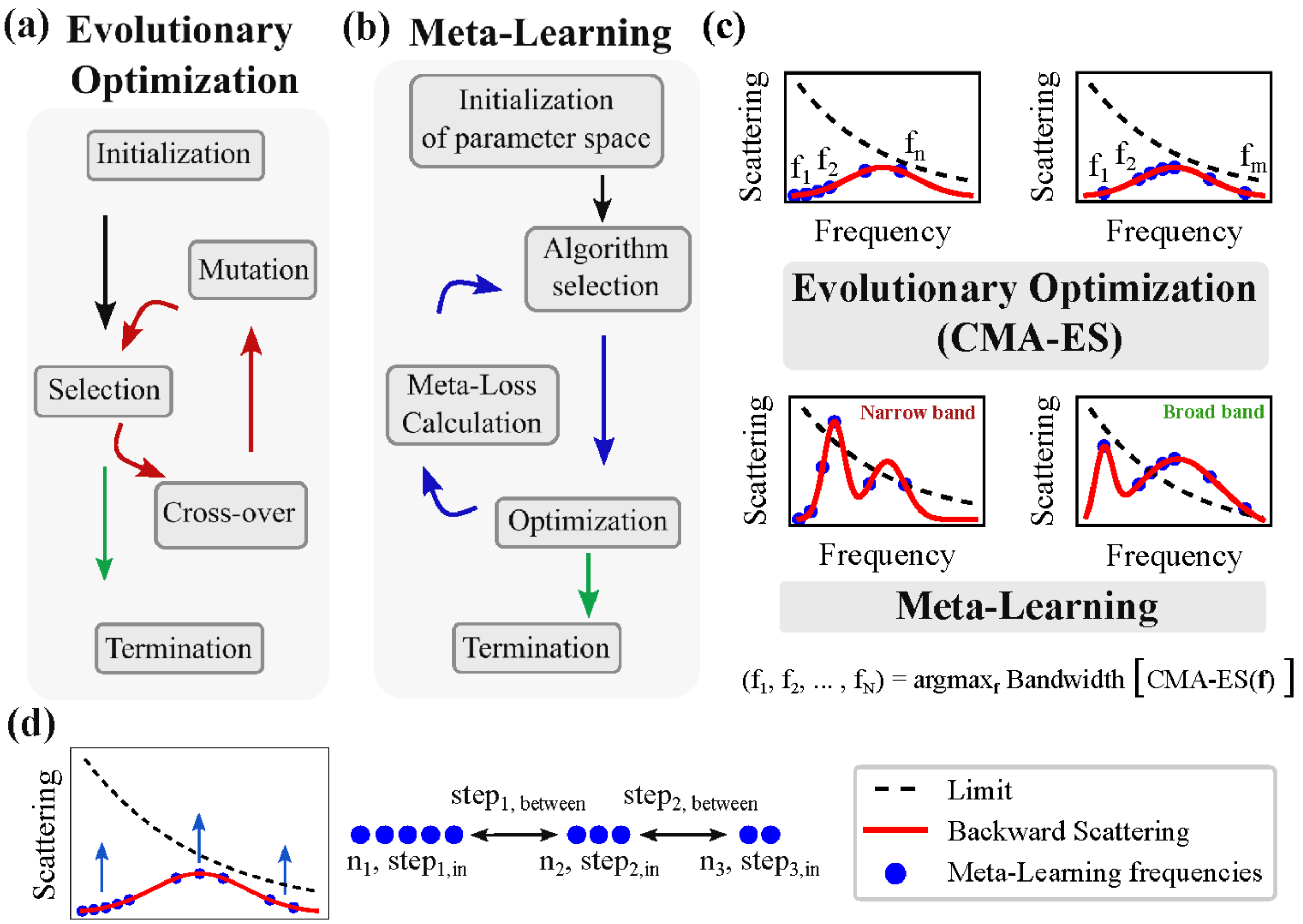

Figure 2. Algorithms for optimizing the superradiant bandwidth. Block diagrams of (a) Genetic optimization and (b) Meta-learning. (c) Manual redistribution of frequency points for assessing the cost function. The adjustment is made manually to improve performance. (d) Meta-learning with Bayesian optimization in several clusters to optimize the spectral sampling.

*Experimental verification*

The meta-learning-based optimization provided arrays whose geometrical parameters appear in the Supplementary Information. The samples are fabricated by chemical etching of copper elements on the dielectric substrate Isola IS680 ($\varepsilon_r$ = 3, tan(δ) = 0.003) with a thickness of 0.127 mm, thus minimizing the impact of the permittivity on the resonant response. Photographs of the samples appear in Fig. 1(c). The measurements were performed in an anechoic chamber. The experimental setup consisted of two closely situated broadband horn antennas NATO IDPH-2018 (Tx and Rx – transmit and receive, certified for 2-18 GHz frequency range), connected to a PNA Vector Network Analyzer N5253b. The backscattering cross-section was quantified using a calibration target (disc) with known parameters[15].

Figures 3(a) demonstrate the backscattering cross-sections (in m$^2$) spectra for three arrays of different size. The superradiant limit is indicated with the black dashed curve. The curve has a hyperbolic shape since the single channel limit depends on the operation wavelength squared ($\lambda^2$). The superradiant bandwidth is defined as a range of frequencies where the red curve is above the black dashed one. The structures possess 2.53 GHz, 4.12 GHz, and 3.99 GHz superradiant bandwidths for 3 x 3, 4 x 4, and 5 x 5 arrays, respectively. It is also worth noting that owing to the meta-learning approach, the backscattering spectra are smooth and almost lack ripples. Only the 5 x 5 array (in the right of Fig.3(a)) demonstrates a notch in the spectrum. In this case, too many multipolar contributions interfere with each other, overcomplicating the optimization procedure and demanding too much computational power.

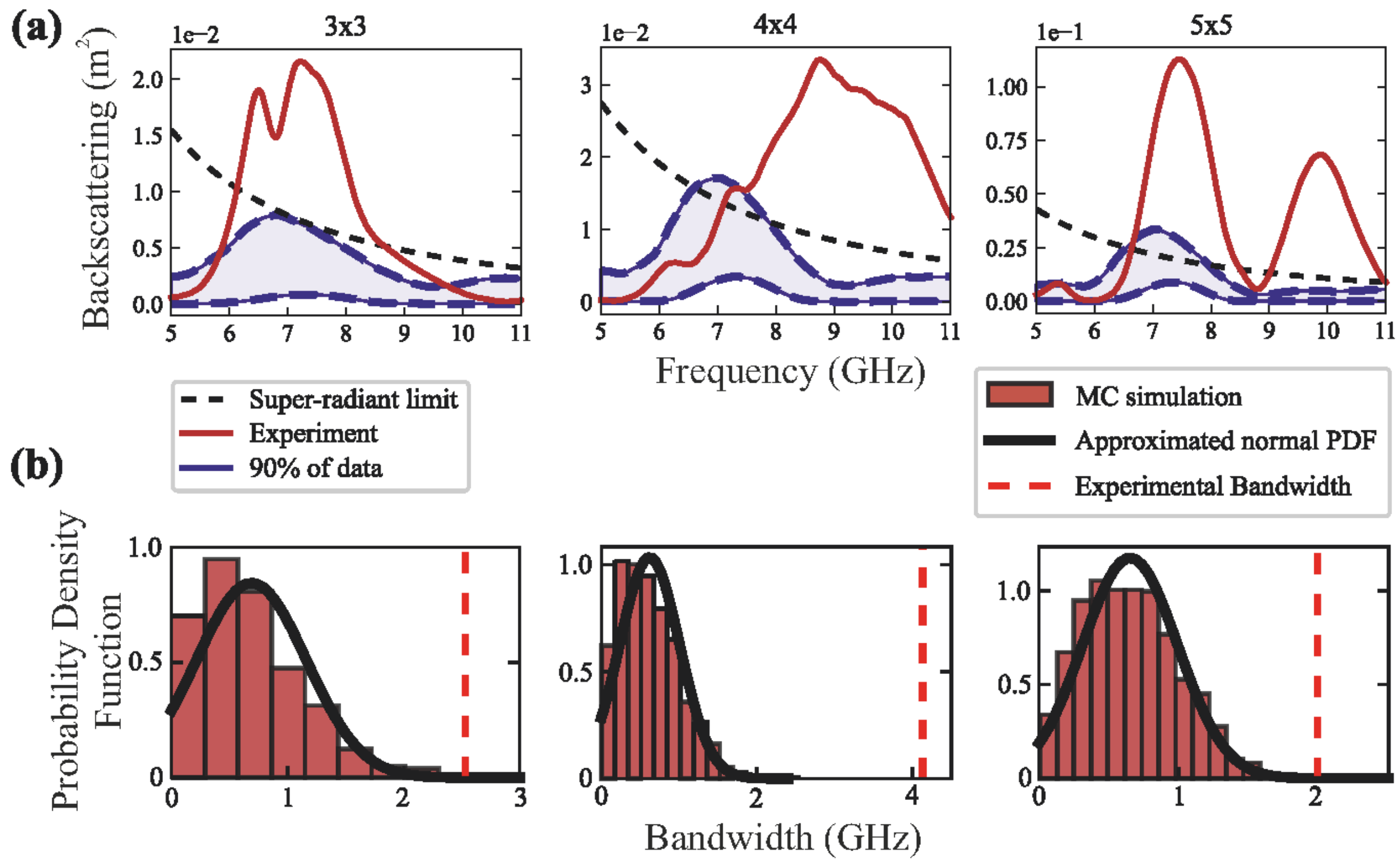

Fig 3. Superscattering spectra, experiment. (a) – red solid line - backscattering spectra for 3 x 3, 4 x 4, and 5 x 5 arrays, respectively. Black-dashed line – superradiant scattering limit. Blue dashed lines - bounds on the performances obtained by randomly generated (non-optimized) structures. (b) – probability density function (PDF) to superradiant bandwidth with random structures (Monte-Carlo simulation). Red-dashed vertical line – the experimentally obtained superradiant bandwidth – far beyond the distribution tail.

*Performance analyses*

A statistical assessment was carried out to understand the rarity of the structures obtained as a result of the optimization. This analysis addresses whether aggressive optimization is needed or similar performances can be obtained by considering many structures with randomly chosen parameters and picking up the most successful realization. For this purpose, a Monte Carlo simulation with 10,000 random structures was collected for the 3 x 3 structure, 5,000 for 4 x 4, and 2,000 for 5 x 5, respectively. The parameters were uniformly distributed within the search space. All the realizations were assessed by their superradiant bandwidth. Panels in Fig 3(b) are histograms, where the vertical axis stands for the probability and the horizontal axis stands for the realized superradiant bandwidth. The red dashed vertical line indicates the bandwidth of experimentally realized structures. The histograms were numerically fitted with normal Gaussian distributions, corresponding to the law of large numbers. Black solid curves are the distributions. It can be clearly seen that the structures obtained by meta-learning are situated significantly away from the distribution, and, in practice, none of the random trials approached their performances. Another result of statistical analysis demonstrates that the probability of accidentally finding structures with such a superradiant bandwidth is less than 0.01 % for the 3 x 3 array. For the 4 x 4, the probability is already less than $10^{-13}$!

*Multipole expansion*

A multipole expansion of the scattering cross-section has been made to reveal the resonance cascading principle governing the electromagnetic performance. Results are shown in Fig. 4. Here, a 3 x 3 structure has been assessed, and six lower-order multipoles were considered for the analysis. Higher-order contributions might be introduced (though having more complex expressions) to obtain a better convergence. For the sake of comparison with the optimization results, the numerical model considers metal strips without a substrate. Thus, the calculations support the experimental data qualitatively. The multipole spectra appear as dashed lines. 3D-scattering diagrams, corresponding to several points along the spectrum, appear in insets. The resonant cascading principle can be identified. Surface currents on the resonators appear in Fig. 4(b), verifying that different parts of the array are responsible for building up the scattering cross section at different frequencies across the spectrum.

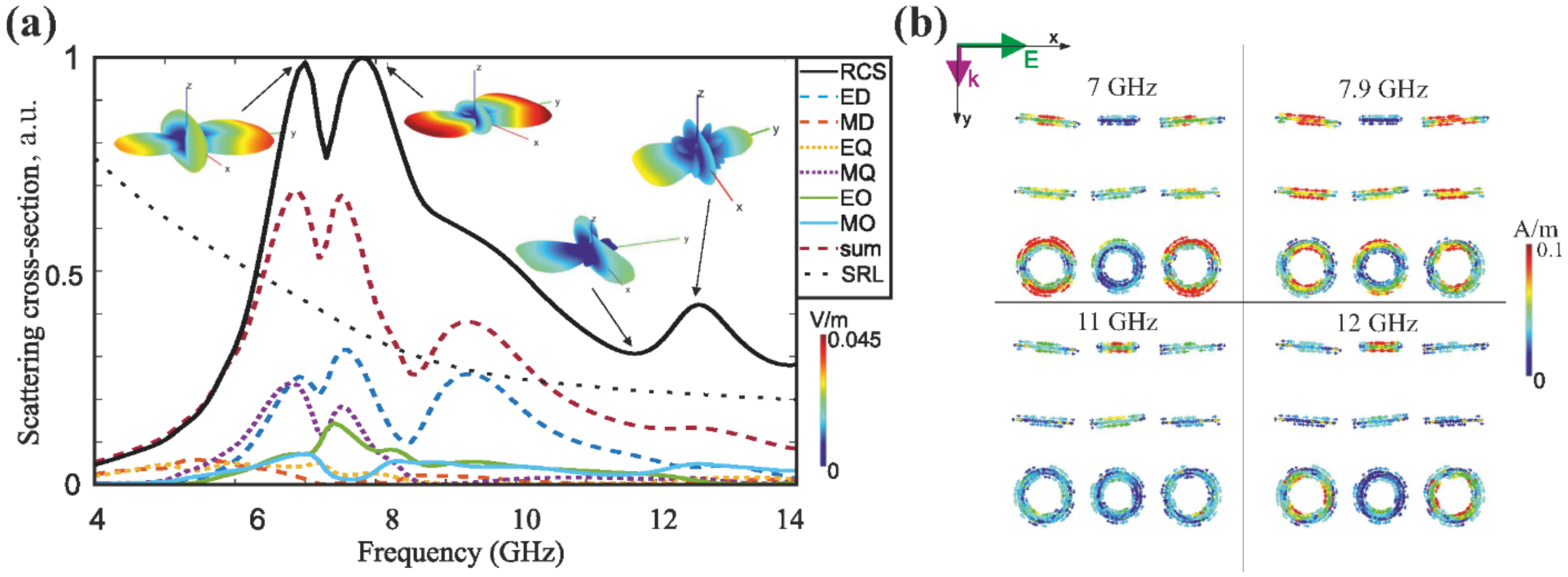

Figure 4. (a) Multipole expansion of the scattering cross-section of 3 x 3 array – numerical results. Excitation – linearly polarized (x-axis) 1 V/m amplitude plane wave propagating along the y-axis. The ED, MD, MD, EQ, MQ, EO, and MO are electric and magnetic dipoles, electric and magnetic quadrupoles, and electric and magnetic octupoles, respectively, black-dashed line - superradiant scattering limit (SRL). In insets, scattering diagrams (far-field electric field amplitude) calculated at 7 GHz, 7.9 GHz, 11.7 GHz, and 12.62 GHz, respectively, are shown. (b) Surface currents, calculated at 7 GHz, 7.9 GHz, 11.7 GHz, and 12.62 GHz, respectively.

## III. CONCLUSION

Conventional electromagnetic wisdom suggests a significant bandwidth degradation with a footprint reduction of scattering structures. This widely believed realm was contested here with the aid of meta-learning-based optimization. Magneto-electric arrays, encompassing resonant wires and split-rings, were optimized towards high backscattering cross-sections over an extended frequency range. The superradiant bandwidth (range of frequencies, where the scattering cross-section prevails over the number of resonators in the array times the single channel limit) has been introduced and optimized. 3 x 3, 4 x 4, and 5 x 5 arrays were explored and demonstrated more than 40 % ultra-wideband superradiant bandwidth. The strength of meta-learning optimization has been shown by comparing the performance of optimized structures versus randomly obtained realizations. It has been shown that over a billion trials must be realized before approaching similar scattering efficiencies.

The superradiant optimization can be related to the so-called radar chaff problem, questioning whether near-field coupling between closely packed resonators can improve the scattering cross-section of structures, given the coupling is neglected. In this context, the current demonstration provides an answer, favoring smartly-optimized structures over randomness.

Highly scatting compact electromagnetic structures with a significant fractional bandwidth have a range of applications in wireless communications, relaxing the need to trade the effective device area for performance.

Acknowledgments:

This work was supported in part by the Department of the Navy, Office of Naval Research Global, under ONRG Award N62909–21–1–2038 and ISF (1115/23).

**Supplementary Information**

The structures parameters:

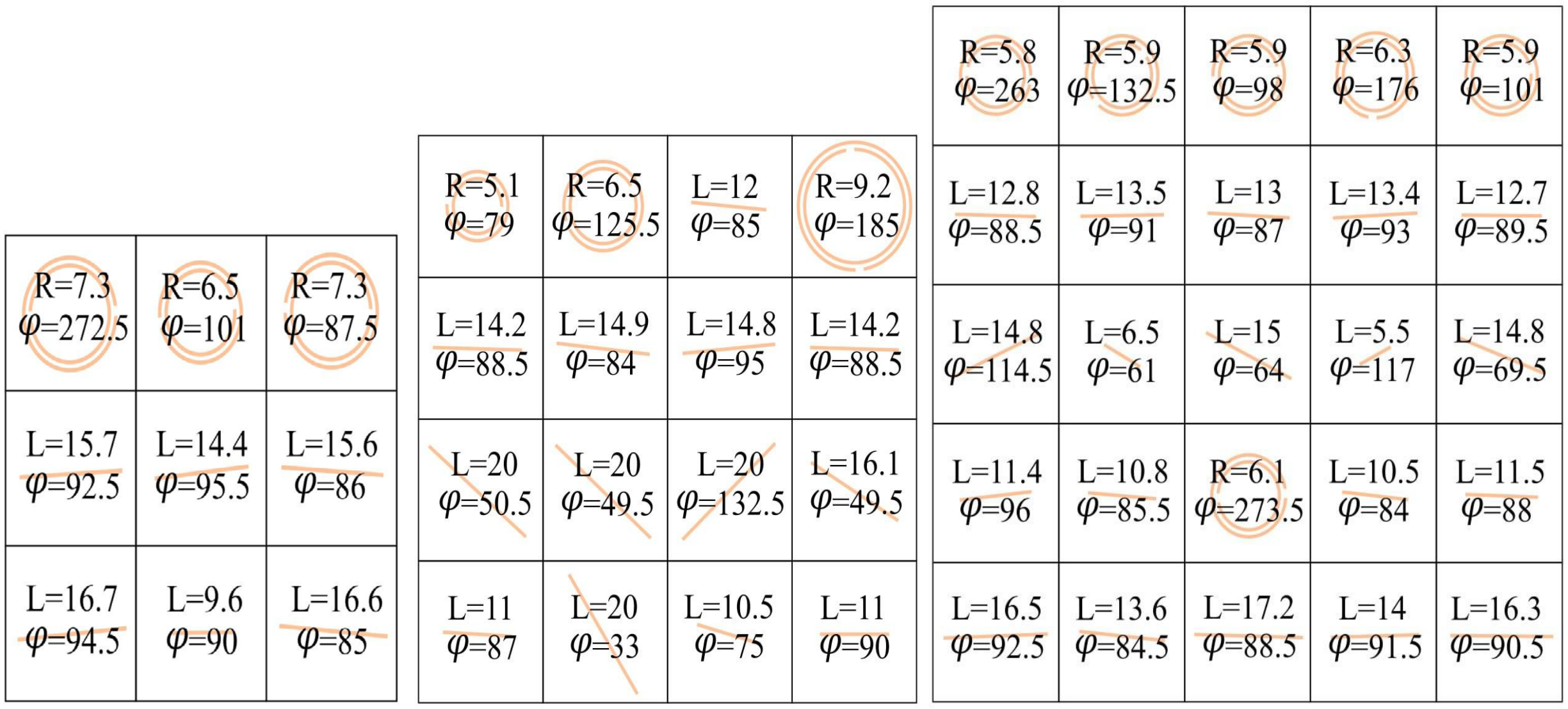